\documentclass[12pt]{article}
\usepackage{amsfonts}

\begin{document}

\title{ Supersymmetric Lorentz invariant deformations of superspaces}

\author {A.A.~Zheltukhin${}^{a,b}$\\
{\normalsize ${}^a$ Kharkov Institute of Physics and Technology, 61108 Kharkov, Ukraine}\\
{\normalsize ${}^{b}$ Institute of Theoretical Physics, University of Stockholm}\\
{\normalsize  SE-10691, AlbaNova, Stockholm, Sweden}}                                            
\date{}

\maketitle

\begin{abstract}
  
 Lorentz invariant supersymmetric deformations of superspaces based on Moyal star product parametrized by Majorana spinor $\lambda_{a}$ and Ramond grassmannian vector $\psi_{m}=-{1\over 2}(\bar\theta\gamma_{m}\lambda)$ in the spinor realization \cite{VZ} are proposed. The map of supergravity background into composite supercoordinates:
$(B^{-1}_{mn}, \Psi^{a}_{m}, C_{ab}) \leftrightarrow (i\psi_{m}\psi_{n}, \psi_{m}\lambda^{a}, \lambda_{a}\lambda_{b})$ valid up to the second order corrections in deformation parameter $h$ and transforming the background dependent Lorentz noninvariant (anti)commutators of supercoordinates into their invariant Moyal brackets is revealed. We found one of the
deformations to depend on 
the axial vector $\psi_{1m}=\frac{1}{2}(\bar\theta\gamma_{m}\gamma_{5}\lambda)$ and to vanish for the $\theta$ components with the same chiralities. The deformations in the (super)twistor picture are discussed.

\end{abstract}

\section{Introduction}

Studying noncommutative  geometry attracts a great interest \cite{Sny, Casal, BFFLS, BrSch, SchNwh, Con, Man, Kon, BaFiShSu, CoDoSch, SeWi, MaSSWes, KLMa, DoN, Sza, FeL,OoVa, KPT, BoGNwh, Sei}. Much attention has been paid to the role of the constant background fields of supergravity - $B_{mn}$, the graviphoton $C_{ab}$ and the gravitino $\Psi^{a}_{m}$ - as the souces of the superspace deformations \cite{SeWi}, \cite{FeL}, \cite{OoVa}, \cite{KPT}, \cite{BoGNwh}, \cite{Sei}.
The presence of the constant background in (anti)commutators of the (super)coordinate operators has stated the problem of the Lorentz symmetry breaking introduced by the deformations. The proposal to overcome this problem by the transition to a twisted Hopf algebra interpretation was recently advanced \cite{ChKNT} and its supersymmetric generalization was developed in \cite{KoSa}, \cite{BZ}, \cite{IhM}. 
Another possibility arises from \cite{UZnc}, where the Hamiltonian and quantum structures of the  twistor-like model \cite{ZUm} of super p-brane embedded in $N=1$ superspace extended by tensor central charge coordinates were studied. The Lorentz covariant supersymmetric non(anti)commutative Dirac bracket relations among the brane (super)coordinates with their r.h.s. parametrized by auxiliary spinor variables were derived there. It hints on a hidden spinor structure associated with the Penrose twistor picture \cite{Penrose, Ferb,Witt, Shir, BC} behind the non(anti)commutativity.
To this end we start here with a spinor extension of the $N=1\, D=4$ superspace $(x_{m},\theta_{a})$ by one commuting Majorana spinor $\lambda_{a}$ and construct Lorentz invariant supersymmetric Poisson and Moyal brackets generating non(anti)commutative relations for  the (super)coordinates. The r.h.s of the $x_m$ brackets among themselves and $x_m$ with $\theta_a$ contain 
 Ramond grassmannian vector $\psi_{m}$ known from the theory of spinning strings and particles \cite{GSW}, \cite{BDZDH}, \cite{Zkk}. The Ramond vector  $\psi_{m}$ appears here in the spinor realization  $\psi_{m}=-{1\over 2}(\bar\theta\gamma_{m}\lambda)$ bilinear in $\lambda_{a},\theta_{a}$ revealed in  \cite{VZ}. The vector $\psi_{m}$ is  associated  with  the spin degrees of freedom in the structure of deformed superspace. 
We revealed a correspondence between the constructed  Lorentz invariant Moyal brackets and the above mentioned (anti)commutators depending of the  constant supergravity  background and string length  $\sqrt{\alpha'}$. This correspondence is schematically illustrated by the map: $B^{-1}_{mn}\leftrightarrow i\psi_{m}\psi_{n},\quad C_{ab}\leftrightarrow \lambda_{a}\lambda_{b},\quad \Psi^{a}_{m}\leftrightarrow \psi_{m}\lambda^{a}$ 
transforming the field dependent (anti)commutators into the Moyal brackets.
 We found  that the map is valid up to the second order corrections in the deformation parameter $h$ and it  works in more sophisticated cases considered below. 
 We studied the null twistor realization of the brackets  and observed the dependence of the non(anti)commutativity effect on the choice of effective variables used to describe the primary degrees of freedom. This observation gives a sudden example of possible couplings between commutative and noncommutative geometries in superspaces.
For the second of the studied  Poisson/Moyal brackets we found a composite grassmannian axial vector $\psi_{1m}={1\over 2}(\bar\theta\gamma_{m}\gamma_{5}\lambda)$ to appear as the deformation measure together with $\psi_{m}$. Moreover, we found the Lorentz invariant brackets for the $\theta$ components with the same  chiralities to be vanising.
 The generalizations of the studied deformations to higher $D=2,3,4 (mod 8)$,  for the case of extended supersymmetries and  for the  presence of additional auxiliary spinors were outlined.

\section{Lorentz invariant splitting of SUSY algebra}

The $D=4\, N=1$ supersymmetry transformations  in the presence of the twistor-like 
Majorana spinor $(\nu_\alpha, \bar\nu_{\dot\alpha})$  are given  by the  relations \cite{UZnc}
\begin{equation}\label{1/8}
\begin{array}{c}
\delta\theta_\alpha=\varepsilon_\alpha,\quad 
\delta x_{\alpha\dot\alpha}=
2i(\varepsilon_{\alpha}\bar\theta_{\dot\alpha}-
\theta_{\alpha}\bar\varepsilon_{\dot\alpha}), \quad \delta\nu_\alpha=0,          
\end{array}
\end{equation}
 and  the correspondent supersymmetric derivatives  $\partial^{\alpha\dot\alpha} \equiv 
\frac{\partial}{\partial x_{\alpha\dot\alpha}}$  and  $D^{\alpha},{\bar D}^{\dot\alpha}$ are  
\begin{equation}\label{3/30}
\begin{array}{c}
D^{\alpha}=\frac{\partial}{\partial\theta_\alpha}-2i\bar\theta_{\dot\alpha}
\partial^{\alpha\dot\alpha}, \quad
{\bar D}^{\dot\alpha}\equiv -(D^{\alpha})^{*}=\frac{\partial}{\partial\bar\theta_{\dot\alpha}}- 2i\theta_{\alpha}\partial^{\alpha\dot\alpha},
 \quad  
[ D^{\alpha},\bar D^{\dot\beta}]=-4i\partial^{\alpha\dot\alpha}.
\end{array}
\end{equation}

The spinor coordinates $ (\nu_{\alpha}, \bar\nu_{\dot\alpha})$ and the light-like vector $\varphi_{\alpha\dot\alpha}=\nu_{\alpha}\bar\nu_{\dot\alpha}$ composed from them may be used  to construct the  Lorentz invariant differential operators 
$ D, \bar D, \partial$
\begin{equation}\label{5/32}
\begin{array}{c}
D= \nu_{\alpha}D^{\alpha},\quad 
\bar D= \bar\nu_{\dot\alpha}\bar {D^{\dot\alpha}},\quad  
\partial=\varphi_{\alpha\dot\alpha} \partial^{\alpha\dot\alpha}
\end{array}
\end{equation}
which  form a supersymmetric subalgebra of the algebra of the invariant derivatives 
\begin{equation}\label{6}
\begin{array}{c} 
[D, \bar D]_{+}=-4i\partial,\quad [D, D]_{+}=[\bar D, \bar D]_{+}=0, \quad [D,\partial]=
 [\bar D,\partial]=[\partial,\partial]=0.
\end{array}
\end{equation}
The superalgebra (\ref{6}) may be  splitted into two invariant and  (anti)commuting subalgebras $(D_{-},\partial)$ and $(D_{+},\partial)$
\begin{equation}\label{8/36}
\begin{array}{c} 
[D_{\pm}, D_{\pm}]_{+}=\mp8i\partial,\quad [D_{+}, D_{-}]_{+}=0, \quad 
[ D_{\pm}, \partial]=[\partial,\partial]=0
\end{array}
\end{equation}
formed  by the supersymmetric derivatives $\partial $ and  $D_{\pm}$
\begin{equation}\label{7}
\begin{array}{c} 
D_{\pm}\equiv D\pm \bar D .
\end{array}
\end{equation}
The addition of the   dilatation operator  $\Delta$
\begin{equation}\label{9/36}
\begin{array}{c}
\Delta=\nu_{\alpha}
\frac{\partial}{\partial\nu_\alpha} + 
\bar\nu_{\dot\alpha}\frac {\partial}{\partial\bar\nu_{\dot\alpha}}
\end{array}
\end{equation}
 changing  the scale of the spinor  $(\nu_\alpha, \bar\nu_{\dot\alpha})$ extends the 
supersubalgebras (\ref{8/36}) 
to the superalgebras formed by the invarint derivatives  $(D_{-},\partial,\Delta )$ and $(D_{+},\partial,\Delta )$
\begin{equation}\label{11/37}
\begin{array}{c} 
[D_{\pm}, D_{\pm}]_{+}=\mp8i\partial,
\quad 
[\Delta,D_{\pm}]= D_{\pm},\quad 
[\Delta,\partial]=2\partial,
\\[0.2cm] 
[D_{+}, D_{-}]_{+}=[ D_{\pm},\partial]=[\partial,\partial]=[\Delta,\Delta]=0.
\end{array}
\end{equation}

 Our proposal is to use the Lorentz invariant supersymmetric differential operators (\ref{11/37}) as building blocks for the construction of Lorentz invariant supersymmetric Poisson and Moyal brackets among the (super)coordinates corresponding to (anti)commutators of the supercoordinate operators in quantum theory.

\section{Supersymmetric Lorentz invariant Poisson bracket}

At first let us study  a simple  example of the Lorentz invariant and supersymmetric Poisson bracket producing  non(anti)commutative relations  among  the superspace coordinates $x_{\alpha\dot\alpha},\theta_{\alpha}, \bar\theta_{\dot\alpha}$. Such a Poisson bracket  may be  constructed from the three differential operators $(D_{-},\partial,\Delta )$ generating the (-)- superalgebra (\ref{11/37})
\begin{equation}\label{13/31}
\begin{array}{c} 
\{ F, G \}= F\,[\, -\frac{i}{4}\,
{\stackrel{\leftarrow}{D}}_{-}
{\stackrel{\rightarrow}{D}}_{-}+ (\stackrel{\leftarrow}{\partial}\stackrel{\rightarrow}
{\Delta} - \stackrel{\leftarrow}{\Delta}\stackrel{\rightarrow}{\partial})\,]\,G ,
\end{array}
\end{equation}
where $ \{  , \}_{P.B.}\equiv \{  , \}$ and $F(x,\theta,\bar\theta,\nu,\bar\nu), G(x,\theta,\bar\theta,\nu,\bar\nu) $ are generalized superfields depending on the superspace coordinates $( x,\theta,\bar\theta)$ and the commuting Weyl spinors $\nu, \bar\nu$. More information on the mathematical definitions used here may be found in \cite{BW},\cite{Zmoy}.

As a result of (\ref{13/31}), the twistor-like coordinates form zero P.B's. among themselves  \begin{equation}\label{17/1}
\{\nu_{\alpha}, \nu_\beta\}=\{\nu_\alpha,\bar\nu_{\dot\beta}\}=
\{\bar\nu_\alpha, \bar\nu_{\dot\beta}\}=0
\end{equation}
and with the Grassmannian spinors $\theta_\alpha, \bar\theta_{\dot\alpha}$ 
\begin{equation}\label{18/2}
\begin{array}{c}
\{\nu_\alpha,\theta_\beta\}=
\{\nu_\alpha, \bar\theta_{\dot\beta}\}=
\{ \bar\nu_{\dot\alpha},\theta_\beta \}=
\{ \bar\nu_{\dot\alpha},\bar\theta_{\dot\beta}\}=0.
\end{array}
\end{equation}
However, they have  non zero P.B's.  with  the space-time coordinates  $x_{\alpha\dot\alpha}$ 
\begin{equation}\label{19/3}
\{ x_{\alpha\dot\alpha}, \nu_\beta \}=\varphi_{\alpha\dot\alpha}\nu_\beta,
\quad
\{ x_{\alpha\dot\alpha}, \bar\nu_{\dot\beta} \}=\varphi_{\alpha\dot\alpha}\bar\nu_{\dot\beta},
\end{equation}
The P.B's. among the super coordinates  $x_{\alpha\dot\alpha}$ and $(\theta_\alpha, \bar\theta_{\dot\alpha})$ are as  follows
\begin{equation}\label{20/7}
\begin{array}{c}
\{x_{\alpha\dot\alpha},x_{\beta\dot\beta}\}=-i\psi_{\alpha\dot\alpha}
\psi_{\beta\dot\beta},
\\[0.2cm]
\{x_{\alpha\dot\alpha},\theta_\beta \}= \frac{i}{2}
\psi_{\alpha\dot\alpha}\nu_{\beta},
\quad
\{x_{\alpha\dot\alpha},\bar\theta_{\dot\beta} \}=-\frac{i}{2}
\psi_{\alpha\dot\alpha}\bar\nu_{\dot\beta},
\\[0.2cm]
\{\theta_{\alpha},\theta_{\beta} \}=\frac{i}{4}\varphi_{\alpha\beta}, 
\quad
\{\theta_\alpha,\bar\theta_{\dot\beta} \}=-\frac{i}{4}\varphi_{\alpha\dot\beta}, 
\quad
\{\bar\theta_{\dot\alpha}, \bar\theta_{\dot\beta}\}=\frac{i}{4}\bar\varphi_{\dot\alpha\dot\beta},
\end{array}
\end{equation}
where  $\psi_{\alpha\dot\alpha}$ is a  Grassmannian vector and $\varphi_{\alpha\beta}, \bar\varphi_{\dot\alpha\dot\beta}$ are composed symmetric spin-tensors 
\begin{equation}\label{21/6}
\psi_{\alpha\dot\alpha}\equiv i(\nu_{\alpha}\bar\theta_{\dot\alpha}- 
\theta_{\alpha}\bar\nu_{\dot\alpha}), \quad
\psi_{\alpha\dot\alpha}\varphi^{\alpha\dot\alpha}=0,
\quad
\varphi_{\alpha\beta}\equiv\nu_{\alpha}\nu_{\beta}, \quad
\bar\varphi_{\dot\alpha\dot\beta}\equiv  \bar\nu_{\dot\alpha}\bar\nu_{\dot\beta},
\quad 
\end{equation} 
with the following transformation rules under the supersymmetry (\ref{1/8}) 
\begin{equation}\label{22/5}
\begin{array}{c}
\delta\varphi_{\alpha\beta}=\delta\bar\varphi_{\dot\alpha\dot\beta}=0, 
\quad
\delta\psi_{\alpha\dot\alpha}=-i(\varepsilon_\alpha \bar\nu_{\dot\alpha}-
\bar\varepsilon_{\dot\alpha}\nu_\alpha).
\end{array}
\end{equation}
The appearance in (\ref{20/7}) of the Ramond  vector $\psi_{\alpha\dot\alpha}$ (\ref{21/6})
associated with the spin degrees of freedom hints on a spin structure behind the coordinate's non(anti)commutativity.
The bilinear spinor representation  for $\psi_{\alpha\dot\alpha}$ (\ref{21/6}) was previosly found  in \cite{VZ} as the general solution of the Dirac constraints $p^{\alpha\dot\alpha}\psi_{\alpha\dot\alpha}=0=p^{\alpha\dot\alpha}p_{\alpha\dot\alpha}$ for massless spinning particle \cite{BDZDH},\cite{Zkk}.  This spinor representation  has established  equivalence between spinning and Brink-Schwarz superparticles. Thus, we find the  desired component Poisson brackets (\ref{17/1}-\ref{20/7}) which are covariant under the Lorentz and supersymmetry transformations.   

The constructed  P.B's. satisfy the graded Jacobi identities having the standard form 
\begin{equation}\label{23/9}
\{\{A,B \},C\} +(-1)^{(b+c)a}\{\{B,C \},A\}
+ (-1)^{c(a+b)}\{\{C,A \},B\}=0,
\end{equation} 
where  $a,b,c =0,1$  denote the Grassmannian  gradings of $A,B$ and  $C$ respectively. 

The  P.B's. among the supercoordinates and the  composite objects $\psi$ and $\varphi$ are 
\begin{equation}\label{26/12}
\begin{array}{c}
\{\psi_{\alpha\dot\alpha},\psi_{\beta\dot\beta}\}=
-i\varphi_{\alpha\dot\alpha}\varphi_{\beta\dot\beta},
\\[0.2cm]
\{x_{\alpha\dot\alpha},\psi_{\beta\dot\beta}\}=\varphi_{\alpha\dot\alpha}
\psi_{\beta\dot\beta}+ \varphi_{\beta\dot\beta}\psi_{\alpha\dot\alpha},
\\[0.2cm]
\{\psi_{\alpha\dot\alpha}, \theta_{\beta}\}=\frac{1}{2}
\varphi_{\alpha\dot\alpha}\nu_{\beta}, \quad
\{\psi_{\alpha\dot\alpha}, \bar\theta_{\dot\beta}\}=-\frac{1}{2}
\varphi_{\alpha\dot\alpha}\bar\nu_{\dot\beta},
\\[0.2cm]
\{x_{\alpha\dot\alpha},\varphi_{\beta\dot\gamma}\}=2\varphi_{\alpha\dot\alpha}\varphi_{\beta\dot\gamma},
\quad 
\{x_{\alpha\dot\alpha},\varphi_{\beta\dot\gamma}\}=2\varphi_{\alpha\dot\alpha}\varphi_{\beta\dot\gamma},
\quad 
\{ x_{\alpha\dot\alpha},\bar\varphi_{\dot\beta\dot\gamma}\}=2\varphi_{\alpha\dot\alpha}\bar\varphi_{\dot\beta\dot\gamma}.
\end{array}
\end{equation}
Using these Poisson brackets together with the  P.B's. (\ref{17/1}-\ref{20/7})  we obtain 
\begin{equation}\label{27/13}
\begin{array}{c}
\{\{\psi_{\alpha\dot\alpha},\psi_{\beta\dot\beta}\},\psi_{\gamma\dot\gamma}\}=0, 
\\[0.2cm]
\{\{\theta_{\alpha},\theta_{\beta}\},\theta_{\gamma}\}=...=\{\{ \bar\theta_{\dot\alpha},\bar\theta_{\dot\beta}\},\bar\theta_{\dot\gamma}\}=0
\end{array}
\end{equation}
proving  the graded Jacobi identity for the $3\psi$ and  $3\theta$ Jacobi cycles
\begin{equation}\label{jc}
 Cycle\{\{\psi_{\alpha\dot\alpha},\psi_{\beta\dot\beta}\},\psi_{\gamma\dot\gamma}\}= Cycle\{\{\theta_{\alpha},\theta_{\beta}\},\theta_{\gamma}\}=...= Cycle\{\{ \bar\theta_{\dot\alpha},\bar\theta_{\dot\beta}\},\bar\theta_{\dot\gamma}\}=0.
\end{equation}
The  vanishing of the  $3x$ Jacobi cycle:
$ Cycle\{\{x_{\alpha\dot\alpha},x_{\beta\dot\beta}\},x_{\gamma\dot\gamma}\}=0$  follows  from the relation 
\begin{equation}\label{31/17}
\{\{x_{\alpha\dot\alpha},x_{\beta\dot\beta}\},x_{\gamma\dot\gamma}\}=2i(\psi_{\alpha\dot\alpha}\psi_{\beta\dot\beta})\varphi_{\gamma\dot\gamma}+ i(\psi_{\alpha\dot\alpha}\varphi_{\beta\dot\beta}-\psi_{\beta\dot\beta}\varphi_{\alpha\dot\alpha})\psi_{\gamma\dot\gamma}.
\end{equation}
The same result are preserved for other Jacobi cycles proving selfconsistency of the introduced  P.B. (\ref{13/31}) that opens a way for the corresponding invariant Moyal bracket.

\section{Lorentz invariant supersymmetric Moyal bracket}

A transition to quantum picture based on the P.B. (\ref{13/31}) may be done using the Weyl-Moyal correspondence establishing one to one correspondence among  quantum field operators and their symbols acting on the commutative space-time. Then the  quantum dynamics encodes itself in the change of usual product of the  Weyl symbols by their star product
\begin{equation}\label{43/38'}
\begin{array}{c} 
F{\star G}=F\, e^{\{ \frac{-ih}{8}\,
[{\stackrel{\leftarrow}{D}}_{-}
{\stackrel{\rightarrow}{D}}_{-}+ (\stackrel{\leftarrow}{\nabla}\stackrel{\rightarrow}
{\Delta} - \stackrel{\leftarrow}{\Delta}\stackrel{\rightarrow}{\nabla})\,]\}}\,G , 
\end{array}
\end{equation}
where $\nabla\equiv4i\partial$ and  $h$ is a quantum deformation parameter associated with the  expansion 
\begin{equation}\label{45/39'}
\begin{array}{c} 
F{\star G}= FG +  (\frac{-ih}{8})\,F\,[{\stackrel{\leftarrow}{D}}_{-}
{\stackrel{\rightarrow}{D}}_{-}+ (\stackrel{\leftarrow}{\nabla}\stackrel{\rightarrow}
{\Delta} - \stackrel{\leftarrow}{\Delta}\stackrel{\rightarrow}{\nabla})\,]\,G 
 \\[0.2cm]
+ \frac{1}{2!}(\frac{-ih}{8})^2\,  
F\,[{\stackrel{\leftarrow}{D}}_{-}
{\stackrel{\rightarrow}{D}}_{-}+(\stackrel{\leftarrow}{\nabla}\stackrel{\rightarrow}
{\Delta}_{-} - \stackrel{\leftarrow}{\Delta}\stackrel{\rightarrow}{\nabla})]^2 \,G 
+ .....
\end{array}
\end{equation}
The power series  expansion in $h$ (\ref{45/39'}) is presented in the arrow ordered form as 
\begin{equation}\label{45/40'}
\begin{array}{c} 
F{\star G}= FG +  (\frac{-ih}{8})\,F\,[{\stackrel{\leftarrow}{D}}_{-}
{\stackrel{\rightarrow}{D}}_{-} +(\stackrel{\leftarrow}{\nabla}
\stackrel{\rightarrow}{\Delta} - \stackrel{\leftarrow}{\Delta}\stackrel{\rightarrow}{\nabla})\,]\,G \\[0.2cm]
+ \frac{1}{2!}(\frac{-ih}{8})^2\,  
F\,[\,  -11 \stackrel{\leftarrow}{\nabla}\stackrel{\rightarrow}{\nabla}+ 3{\stackrel{\leftarrow}{D}}_{-}\stackrel{\longleftrightarrow}{\nabla}{\stackrel{\rightarrow}{D}}_{-}
+
2( \stackrel{\leftarrow}{\nabla}{\stackrel{\leftarrow}{D}}_{-}{\stackrel{\rightarrow}{D}}_{-}\stackrel{\rightarrow}{\Delta} -  \stackrel{\leftarrow}{\Delta }{\stackrel{\leftarrow}{D}}_{-}{\stackrel{\rightarrow}{D}}_{-}\stackrel{\rightarrow}{\nabla}) 
 \\[0.2cm]
- 3( \stackrel{\leftarrow}{\Delta}\stackrel{\leftarrow}{\nabla}\stackrel{\rightarrow}{\nabla}
+\stackrel{\leftarrow}{\nabla}\stackrel{\rightarrow}{\nabla}\stackrel{\rightarrow}{\Delta})
+\stackrel{\leftarrow}{\nabla}^2({\stackrel{\rightarrow}{\Delta}}^2 +2\stackrel{\rightarrow}{\Delta}) + {(\stackrel{\leftarrow}{\Delta}}^2 + 
2\stackrel{\leftarrow}{\Delta}){\stackrel{\rightarrow}{\nabla}}^2
- 2\stackrel{\leftarrow}{\Delta}\stackrel{\leftarrow}{\nabla}\stackrel{\rightarrow}{\nabla}\stackrel{\rightarrow}{\Delta}
 \,]\,G +.....,
\end{array}
\end{equation}
where we omit the higher order terms in $h$.
Using the  expansion (\ref{45/40'})  we find the second order corrections to be vanishing for the following $\star$-products of the supercoordinates:
\begin{equation}\label{47/42}
\begin{array}{c} 
x_{\alpha\dot\alpha}\star \nu_\beta = x_{\alpha\dot\alpha}\nu_{\beta} +
\frac{h}{2}\varphi_{\alpha\dot\alpha} \nu_{\beta} + {\cal O}(h^3),
\quad
x_{\alpha\dot\alpha}\star\bar\nu_{\dot\beta}=x_{\alpha\dot\alpha}\bar\nu_{\dot\beta}
+\frac{h}{2}\varphi_{\alpha\dot\alpha}\bar\nu_{\dot\beta} + {\cal O}(h^3),
\\[0.2cm]
x_{\alpha\dot\alpha}\star \theta_\beta = x_{\alpha\dot\alpha}\theta_{\beta} +
\frac{ih}{4}
\psi_{\alpha\dot\alpha}\nu_{\beta}+  {\cal O}(h^3),
\quad
x_{\alpha\dot\alpha}\star\bar\theta_{\dot\beta}=x_{\alpha\dot\alpha}\bar\theta_{\dot\beta}
-\frac{ih}{4}
\psi_{\alpha\dot\alpha}\bar\nu_{\dot\beta}+ {\cal O}(h^3),
\\[0.2cm]
\theta_{\alpha}\star \theta_{\beta} =\theta_{\alpha}\theta_{\beta}  +\frac{ih}{8}\varphi_{\alpha\beta}+
 {\cal O}(h^3), \quad
\theta_\alpha\star\bar\theta_{\dot\beta} =\theta_\alpha\bar\theta_{\dot\beta} -\frac{ih}{8}\varphi_{\alpha\dot\beta}+ {\cal O}(h^3), 
\\[0.2cm]
\bar\theta_{\dot\alpha}\star \bar\theta_{\dot\beta}=\bar\theta_{\dot\alpha}\bar\theta_{\dot\beta} +\frac{ih}{8}{\bar\varphi}_{\dot\alpha\dot\beta}+ {\cal O}(h^3).
\end{array}
\end{equation}   
Moreover, the star products of the Majorana spinor components  $(\nu_{\alpha},\bar\nu_{\dot\alpha})$ coincide  with  their  usual products in all orders in $h$. We assume that the higher order corrections in the star products  (\ref {47/42}) can be also equal  zero. On the contrary, the second order corrections in the star products of the  $x_{\alpha\dot\alpha}$ components are  nonzero
\begin{equation}\label{47/43'}
x_{\alpha\dot\alpha}\star x_{\beta\dot\beta}=x_{\alpha\dot\alpha}x_{\beta\dot\beta}-\frac{ih}{2}\psi_{\alpha\dot\alpha}\psi_{\beta\dot\beta}- \frac{11}{2!}\frac{h^2}{4}\varphi_{\alpha\dot\alpha}\varphi_{\beta\dot\beta}+ {\cal O}(h^3),
\end{equation}
but their contributions in the corresponding  Moyal brackets are zero, because of the  commutativity $\varphi_{\alpha\dot\alpha}\varphi_{\beta\dot\beta}=\varphi_{\beta\dot\beta}\varphi_{\alpha\dot\alpha}$. 
Consequently, the second  order corections in the Lorentz invariant and supersymmetric Moyal brackets (\ref{47/42}-\ref{47/43'}) are equal  to  zero 
\begin{equation}\label{48/43}
\begin{array}{c} 
[x_{\alpha\dot\alpha},x_{\beta\dot\beta}]_{\star}\equiv x_{\alpha\dot\alpha}\star x_{\beta\dot\beta}- x_{\beta\dot\beta}\star x_{\alpha\dot\alpha}
=-ih\psi_{\alpha\dot\alpha}\psi_{\beta\dot\beta}+ {\cal O}(h^3),
\\[0.2cm]
[x_{\alpha\dot\alpha},\nu_\beta]_{\star}= h\varphi_{\alpha\dot\alpha}\nu_{\beta}+ {\cal O}(h^3),
\quad
[x_{\alpha\dot\alpha},\bar\nu_{\dot\beta}]_{\star}=h\varphi_{\alpha\dot\alpha}\bar\nu_{\dot\beta}+ {\cal O}(h^3),
\\[0.2cm]
[x_{\alpha\dot\alpha},\theta_\beta]_{\star}= \frac{ih}{2}
\psi_{\alpha\dot\alpha}\nu_{\beta}+ {\cal O}(h^3),
\quad
[x_{\alpha\dot\alpha},\bar\theta_{\dot\beta}]_{\star}=-\frac{ih}{2}
\psi_{\alpha\dot\alpha}\bar\nu_{\dot\beta}+  {\cal O}(h^3),
\\[0.2cm]
[\theta_{\alpha},\theta_{\beta} ]_{\star+}=\frac{ih}{4}\varphi_{\alpha\beta}+ {\cal O}(h^3), 
\quad
[\theta_\alpha,\bar\theta_{\dot\beta}]_{\star+}=-\frac{ih}{4}\varphi_{\alpha\dot\beta}+{\cal O}(h^3), 
\quad
[\bar\theta_{\dot\alpha}, \bar\theta_{\dot\beta}]_{\star+}=\frac{ih}{4}\bar\varphi_{\dot\alpha\dot\beta}+
 {\cal O}(h^3).
\end{array}
\end{equation} 
The Moyal brackets generated by the P.B's. (\ref{17/1}-\ref{20/7}) replace the (anti)commutators of the coordinate operators used in the standard quantum picture.

\section{Brackets and twistors}

The unification of the  Weyl spinors  $\nu_{\alpha},\bar\nu_{\dot\alpha}$ with the spinors  $\omega^{\alpha},\bar\omega^{\dot\alpha}$ defined as 
\begin{equation}\label{49/44}
\omega_{\alpha}=x_{\alpha\dot\alpha}\bar\nu^{\dot\alpha},\quad 
\bar\omega_{\dot\alpha}=x_{\alpha\dot\alpha}\nu^{\alpha}
\end{equation} 
yields the null  twistor  $Z^{\cal A}=(i\omega^{\alpha}, \bar\nu_{\dot\alpha})$ and its  complex conjugate $\bar Z_{\cal A}=(\nu_{\alpha},-i\bar\omega^{\dot\alpha})$ connected  by the condition  $Z^{\cal A}\bar Z_{\cal A}=0 $ \cite{Penrose}. The Eqs. (\ref{17/1}) and (\ref{19/3}) result in the P.B. commutativity among the twistor components $\omega_{\alpha}$ and $\nu_{\alpha},\bar\nu_{\dot\alpha} $
\begin{equation}\label{50/44}
\{\omega_{\alpha}, \nu_{\beta}\}=\{\omega_{\alpha}, \bar\nu_{\dot\beta}\}=\{\bar\omega_{\dot\alpha},\nu_{\beta}\}=\{\bar\omega_{\dot\alpha}, \bar\nu_{\dot\beta}\}=0,
\end{equation}
 because of the orthogonality conditions 
\begin{equation}\label{51/44}
\varphi_{\alpha\dot\alpha}\nu^{\alpha}=\varphi_{\alpha\dot\alpha}\bar\nu^{\dot\alpha}=0.
\end{equation} 
The P.B's. among the components of the Majorana spinor $(\omega_{\alpha},\bar\omega_{\dot\alpha})$ with  the same chirality
\begin{equation}\label{55}
\{\omega_{\alpha},\omega_{\beta}\}=i{\bar\eta}^2\varphi_{\alpha\beta}\equiv0,
\quad 
\{\bar\omega_{\dot\alpha},\bar\omega_{\dot\beta}\}=i{\eta}^2\bar\varphi_{\dot\alpha\dot\beta}\equiv0
\end{equation}
 become zero, because ${\eta}^2={\bar\eta}^2=0$,
where $\eta$ and $\bar\eta$ are grassmannian scalars defined  by 
\begin{equation}\label{53/46}
 \eta\equiv\theta_{\alpha}\nu^{\alpha}, \quad \bar\eta\equiv\bar\theta_{\dot\alpha}\bar\nu^{\dot\alpha}.
\end{equation} 
These anticommuting scalars have zero P.B's. between themselves, with $\nu,\omega,\theta$
\begin{equation}\label{54}
\{\eta, \nu_{\alpha}\} =\{\eta, \omega_{\alpha} \}=\{\eta,\theta_\alpha\}=0
\end{equation}
and with  $\bar\nu,\bar\omega,\bar\theta$.
Single nonzero of the Poisson brackets among the null twistor $Z^{\cal A}=(i\omega^{\alpha}, \bar\nu_{\dot\alpha})$, $\bar Z_{\cal A}=(\nu_{\alpha}, -i\bar\omega^{\dot\alpha})$  components is
\begin{equation}\label{56}
\{\omega_{\alpha},\bar\omega_{\dot\beta}\}=i\eta\bar\eta\varphi_{\alpha\dot\beta}.
\end{equation}
 and it may be written down in the equivalent form 
$
\{\omega_{\alpha},{\bar\omega}_{\dot\beta}\}=
8\eta\bar\eta\{\theta_{\alpha},\bar\theta_{\dot\beta}\}
$ 
showing proportionality of the $(\omega,\bar\omega)$ noncommutativity  tothe $(\theta,\bar\theta)$ nonanticommutativity. It fixes s correlation of the twistor 
deformation with supersymmetry encoded in the Lorentz invariant P.B. (\ref{13/31}). 
This correlation manifests its under reduction of the original superspace to the null supertwistor subspace  formed by  $ Z^{\tilde{\cal A}}, {\bar Z}_{\tilde{\cal A}}$ connected by the 
 relation: $ Z^{\tilde{\cal A}}{\bar Z}_{\tilde{\cal A}}=0 $ \cite{Ferb}. The null supertwistors 
are formed by the  triads 
 $Z^{\tilde{\cal A}}=(iq^{\alpha}, \bar\nu_{\dot\alpha},2\bar\eta),{\bar Z}_{\tilde{\cal A}}=(\nu_{\alpha}, -i{\bar q}^{\dot\alpha},2\eta)$, where $q_{\alpha}=\omega_{\alpha}-2i \bar\eta\theta_{\alpha}$, whose supersubspace is  closed under the supersymmetry transformations. 
Because of this reduction we find the counterpart of the P.B. ({\ref{56}}) to vanish 
 \begin{equation}\label{56/1}
\{q_{\alpha},{\bar q}_{\dot\beta}\}=0
\end{equation}
 together with any other P.B's. among the components of $ Z^{\tilde{\cal A}}, {\bar Z}_{\tilde{\cal A}}$.  It means that the supersubspace of null supertwistors $ Z^{\tilde{\cal A}}, {\bar Z}_{\tilde{\cal A}}$ is inert under the deformation associated with the P.B. (\ref{13/31}). This  effect is a consequence of nonlocal  connection ({\ref{49/44}) between the coordinates and twistors, because the relation ({\ref{49/44}) is invariant under the shifts:
 $ x_{\alpha\dot\alpha} \rightarrow x_{\alpha\dot\alpha}+ s\nu_{\alpha}\bar\nu_{\dot\alpha}$. This shifts map light-like lines into points and wash off the noncommutativity effect originating from the uncertainty relations for the $x_m$ components on the Plank scale.
This observation gives an example of couplings between commutative and noncommutative geometries in the presence of supersymmetry.
So, we observe interesting couplings of twistor structure with supersymmetry, Lorentz invariance and Poisson structure which shed light on general structure of non(anti)commutative superspaces.

\section{The Lorentz invariant bracket in higher dimensions}

The passage to the Majorana representation in the Poisson brackets (\ref{17/1}-\ref{20/7})
\begin{equation}\label{58}
\nu_{a}={\nu_\alpha\choose \bar\nu ^{\dot\alpha}},
\quad
\theta_{a}={\theta _\alpha\choose \bar\theta^{\dot\alpha}},
\quad
C^{ab}=\left(\begin{array}{cc} \varepsilon^{\alpha\beta}&0\\
0&\bar\varepsilon_{\dot\alpha\dot\beta}
 \end{array}\right),
\quad
\chi^{a}=C^{ab}\chi_{b},                                                       
\end{equation}
where  $C^{ab}$ is the charge conjugation matrix, presents them in the form suitable for the generalization to higher dimensions  
\begin{equation}\label{59}
\{\nu_{a},\nu_{b}\}=0, \quad \{\theta_{a},\nu_{b}\}=0, 
\quad
\{x_{m},\nu_{a}\}=\varphi_{m}\nu_{a}.
\end{equation}
The real vectors $x_{m}$ and $\varphi_{m}$ in (\ref{59}) are defined by the relations \cite{BW}
\begin{equation}\label{60}
\begin{array}{c}
x_{m}=-{1\over 2}(\tilde{\sigma}_{m})^{\dot\alpha\beta}x_{\beta\dot\alpha},
\quad
x_{\alpha\dot\beta}=(\sigma^{m})_{\alpha \dot\beta}{x_m}, 
\\[0.2cm]
\varphi_{m}=-{1\over 2}(\tilde{\sigma}_{m})^{\dot\alpha\beta}\varphi_{\beta\dot\alpha}\equiv
{1\over 4}(\bar\nu\gamma_{m}\nu),
\end{array}
\end{equation}
where   $\gamma_{m}$ are the Dirac matrices in the  Majorana representation.

To rewrite the  rest of the P.B's. in the  Majorana representation it is  convenient to change the  Majorana spinor  $\nu_{a}$  by other Majorana spinor $\lambda_{a}$  
\begin{equation}\label{61}
\lambda_{a}={\lambda_{\alpha}\choose \bar\lambda^{\dot\alpha}}\equiv(\gamma_{5}\nu)_{a},
\quad 
(\gamma_{5})_a{}^b=\left(\begin{array}{cc} -i\delta_\alpha^\beta&0\\
0&i\delta^{\dot\alpha}_{\dot\beta} \end{array}\right)
\end{equation}
 preserving  the form of the P.B's. (\ref{59}). In terms of the real  Majorana spinor $\lambda_{a}$  and  the composed  vectors  $\varphi_{m}$ and  $\psi_{m}$ 
\begin{equation}\label{62}
\varphi_{m}={1\over 4}(\bar\lambda\gamma_{m}\lambda),
\quad 
\psi_{m}=-{1\over 2}(\tilde{\sigma}_{m})^{\dot\alpha\alpha}\psi_{\alpha\dot\alpha}\equiv
-{1\over 2}(\bar\theta\gamma_{m}\lambda)
\end{equation}
the P.B's. (\ref{17/1}-\ref{20/7}) of the primordial coordinates $x_{m},\theta_{a},\lambda_{a}$ are presented as follow
\begin{equation}\label{63}
\begin{array}{c}
\{\lambda_{a},\lambda_{b}\}=0,
\quad
\{\theta_{a},\lambda_{b}\}=0,
\quad
\{x_{m},\lambda_{a}\}=\varphi_{m}\lambda_{a},
\\[0.2cm] 
\{x_{m},x_{n}\}=-i\psi_{n}\psi_{m},
\quad
\{x_{m},\theta_{a} \}= -\frac{1}{2}\psi_{m}\lambda_{a},
\quad
\{\theta_{a},\theta_{b} \}=-\frac{i}{4}\lambda_{a}\lambda_{b}. 
\end{array}
\end{equation}

The P.B's. of the composite vectors $\psi_{m}$  and  $\varphi_{m}$ (\ref{62}) among themselves  and  with the primordial coordinates take  the  form
 \begin{equation}\label{64}
\begin{array}{c}
\{x_{m},\psi_{n}\}=\varphi_{m}\psi_{n} + \varphi_{n}\psi_{m},
\quad
\{\psi_{m},\theta_{b} \}= {i\over 2}\varphi_{m}\lambda_{b},
\quad
\{\psi_{m},\lambda_{a} \}=0,
\\[0.2cm]
\{\psi_{m},\psi_{n}\}=-i\varphi_{m}\varphi_{n}, 
\quad
\{\psi_{m},\varphi_{n}\}=0
\end{array}
\end{equation}
and respectively 
\begin{equation}\label{65}
\begin{array}{c}
\{x_{m},\varphi_{n}\}=2\varphi_{m}\varphi_{n},
\quad
\{\theta_{a},\varphi_{m}\}= \{\lambda_{a},\varphi_{m}\}=\{\varphi_{m},\varphi_{n} \}=0.
\end{array}
\end{equation}
The P.B's. (\ref{63}-\ref{65}) originally derived for $ D=4$ are valid in $D$-dimensional space with $D=2,3,4 (mod 8)$, where the Majorana spinors exist. This procedure restores the vector form of the Moyal brackets (\ref{48/43}) in the higher dimensions.

\section{Other supersymmetric Lorentz invariant brackets}

Using the Majorana spinor $\nu_a$ one can constuct one more supersymmetric and Lorentz  invariant Poisson bracket in the addition to the P.B. (\ref{13/31}) which is  given by 
\begin{equation}\label{66/3}
\begin{array}{c} 
\{ F, G \}= F\,[\, \frac{i}{4}\,(
\stackrel{\leftarrow}{D}\stackrel{\rightarrow}{\bar D}+
\stackrel{\leftarrow}{\bar D}\stackrel{\rightarrow}{D}) + 
\frac{1}{2}(\stackrel{\leftarrow}{\partial}\stackrel{\rightarrow}
{\Delta} - \stackrel{\leftarrow}{\Delta}\stackrel{\rightarrow}{\partial})\,]\,G
\end{array}
\end{equation}
and yields different  invariant  Poisson brackets for the supercoordinates $x$ and $\theta$
 \begin{equation}\label{66/4}
\begin{array}{c}
\{x_{\alpha\dot\alpha},x_{\beta\dot\beta}\}=
-i(\varphi_{\alpha\dot\beta}{\bar\theta}_{\dot\alpha}\theta_\beta -
\varphi_{\beta\dot\alpha}{\bar\theta}_{\dot\beta}\theta_{\alpha}),
\\[0.2cm]
\{x_{\alpha\dot\alpha},\theta_\beta \}= \frac{1}{2}\varphi_{\beta\dot\alpha}\theta_{\alpha},
\quad
\{x_{\alpha\dot\alpha},\bar\theta_{\dot\beta} \}=\frac{1}{2}
\varphi_{\alpha\dot\beta}\bar\theta_{\dot\alpha},
\\[0.2cm]
\{\theta_{\alpha},\theta_{\beta} \}=\{\bar\theta_{\dot\alpha}, \bar\theta_{\dot\beta}\}=0, 
\quad
\{\theta_\alpha,\bar\theta_{\dot\beta} \}=-\frac{i}{4}\varphi_{\alpha\dot\beta}
 \end{array}
\end{equation}
 We see that the new deformation (\ref{66/3}) generates the zero P.B's. for  the  $\theta_{a}$ components  with the same chirality in contrast to the deformation  (\ref{13/31}).
The P.B's. (\ref{66/4}) are added  by 
\begin{equation}\label{66/4'}
\begin{array}{c}
\{\nu_{\alpha}, \nu_\beta\}=\{\nu_\alpha,\bar\nu_{\dot\beta}\}=
\{\bar\nu_\alpha, \bar\nu_{\dot\beta}\}=0,
\\[0.2cm]
\{\nu_\alpha,\theta_\beta\}=
\{\nu_\alpha, \bar\theta_{\dot\beta}\}=
\{ \bar\nu_{\dot\alpha},\theta_\beta \}=
\{ \bar\nu_{\dot\alpha},\bar\theta_{\dot\beta}\}=0,
\\[0.2cm]
\{ x_{\alpha\dot\alpha}, \nu_\beta \}= \frac{1}{2}\varphi_{\alpha\dot\alpha}\nu_\beta,
\quad
\{ x_{\alpha\dot\alpha}, \bar\nu_{\dot\beta} \}= \frac{1}{2}\varphi_{\alpha\dot\alpha}\bar\nu_{\dot\beta},
\end{array}
\end{equation}
  The  P.B. (\ref{66/3}) satisfies the Jacobi identities and produces  the corresponding Moyal bracket  
\begin{equation}\label{66'}
\begin{array}{c} 
F{\star G}=F\, e^{\{ \frac{ih}{8}\,[\stackrel{\leftarrow}{D}\stackrel{\rightarrow}{\bar D}+
\stackrel{\leftarrow}{\bar D}\stackrel{\rightarrow}{D}) 
- \frac{1}{2}(\stackrel{\leftarrow}{\nabla}\stackrel{\rightarrow}
{\Delta} - \stackrel{\leftarrow}{\Delta}\stackrel{\rightarrow}{\nabla})\,]\}}\,G , 
\end{array}
\end{equation}
where $\nabla\equiv4i\partial$ and  $h$ is a quantum deformation parameter.

Using the conversion formulae from Sect. 6 gives the vector form for the P.B's. (\ref{66/4})
\begin{equation}\label{66/5'}
\begin{array}{c}
\{x_{m},x_{n}\}=-\frac{i}{4}(\chi_{m}\bar\chi_{n}-\chi_{n}\bar\chi_{m}),
\\[0.2cm]
\{x_{m},\theta_{\beta}\}=-\frac{1}{4}{\bar\chi}_{m}\nu_{\beta}, \,\, \,
\{x_{m},{\bar\theta}_{\dot\beta}\}=-\frac{1}{4}\chi_{m}{\bar\nu}_{\dot\beta},
\\[0.2cm]
\{\theta_{a},\theta_{b}\}=-\frac{i}{8}(\nu^{(+)}_{a}\nu^{(-)}_{b}+\nu^{(+)}_{b}\nu^{(-)}_{a}),
\end{array}
\end{equation}
where we introduced  the complex Grasssmannian vector $\chi_{m}$  with the real and imaginary parts  presented by $\psi_{1m}, \psi_{2m}$  and the chiral components  $\theta^{(\pm)}$ and $\nu^{(\pm)}$ 
\begin{equation}\label{66/5''}
\begin{array}{c}
\chi_{m}\equiv
(\nu\sigma_{m}\bar\theta)\equiv-\bar\nu\gamma_{m}\frac{1+i\gamma_{5}}{2}\theta\equiv \psi_{1m}+ i\psi_{m},
\\[0.2cm]
\bar\chi_{m}\equiv(\chi_{m})^{*}= -\bar\nu\gamma_{m}\frac{1-i\gamma_{5}}{2}\theta,
\quad
\psi_{1m}\equiv -\frac{1}{2}(\bar\theta\gamma_{m}\nu),
\quad
\psi_{m}\equiv -\frac{1}{2}(\bar\theta\gamma_{m}\gamma_{5}\nu),
\\[0.2cm]
\theta^{(\pm)}\equiv\frac{1}{2}
(1 \pm i\gamma_{5})\theta, \,\,\,\,
\nu^{(\pm)}\equiv\frac{1}{2}(1 \pm  i\gamma_{5})\nu .
\end{array}
\end{equation}

Then the  P.B's. (\ref{66/5'}) are presented in the form directly generalizing the P.B's.  (\ref{63})
\begin{equation}\label{66/5'''}
\begin{array}{c}
\{x_{m},x_{n}\}=-\frac{i}{2}(\psi_{1m}\psi_{1n} + \psi_{m}\psi_{n}),
\\[0.2cm]
\{x_{m},\theta_{a}\}=-\frac{1}{4}(\psi_{1m}\nu_{a} + \psi_{m}\lambda_{a}),
\\[0.2cm]
\{\theta_{a},\theta_{b}\}=-\frac{i}{8}(\nu_{a}\nu_{b}+\lambda_{a}\lambda_{b}),
\end{array}
\end{equation}
where $\lambda_{a}\equiv(\gamma_{5}\nu)_{a}$ as in (\ref{61}).
Comparing (\ref{66/5'''}) with (\ref{63}) we observe that the change of the P.B. (\ref{13/31}) by 
(\ref{66/3}) is equivalent to the complexification of the real Grassmannian vector $\psi_{m}$ (\ref{62}) accompanied by the appearance of the spinors $\nu_{a}$ and 
$(\gamma_{5}\nu)_{a}$ in the r.h.s. of  (\ref{66/5'''}).

The P.B. (\ref{66/3}) and  respectively the Moyal bracket (\ref{66'}) may be generalized to the case of 
extended supersymmetries with  $N>1$. The corresponding  P.B. may be chosen as
\begin{equation}\label{66/5}
\begin{array}{c} 
\{ F, G \}= F\,[\, \frac{i}{4}\,(
\stackrel{\leftarrow}{D_{i}}\stackrel{\rightarrow}{\bar D^{i}}+
\stackrel{\leftarrow}{\bar D^{i}}\stackrel{\rightarrow}{D_{i}}) + 
\frac{1}{2}(\stackrel{\leftarrow}{\partial}\stackrel{\rightarrow}
{\Delta} - \stackrel{\leftarrow}{\Delta}\stackrel{\rightarrow}{\partial})\,]\,G,
\end{array}
\end{equation}
where $D_{i}\equiv\nu_\alpha D^{\alpha}_{i}$ and ${\bar D}^{i}\equiv{\bar\nu}_{\dot\alpha}
{\bar D}^{\dot\alpha i}$ \, with  i=1,2,..,$N$. 
The P.B's. (\ref{66/5}) generate  the following brackets  for the primordial  (super)coordinates 
\begin{equation}\label{66/6}
\begin{array}{c}
\{x_{\alpha\dot\alpha},x_{\beta\dot\beta}\}=
-i(\varphi_{\alpha\dot\beta}{\bar\theta}_{\dot\alpha i}\theta^{i}_{\beta} -
\varphi_{\dot\alpha\beta}{\bar\theta}_{\dot\beta i}\theta^{i}_{\alpha}),
\\[0.2cm]
\{x_{\alpha\dot\alpha},\theta^{i}_\beta \}= \frac{1}{2}\varphi_{\dot\alpha\beta}\theta^{i}_{\alpha},
\quad
\{x_{\alpha\dot\alpha},\bar\theta_{\dot\beta i} \}=\frac{1}{2}
\varphi_{\alpha\dot\beta}\bar\theta_{\dot\alpha i},
\\[0.2cm]
\{\theta^{i}_{\alpha},\theta^{k}_{\beta} \}=\{\bar\theta_{\dot\alpha i}, \bar\theta_{\dot\beta k}\}=0, 
\quad
\{\theta^{i}_{\alpha},{\bar\theta}_{\dot\beta k } \}=\frac{i}{4}\varphi_{\alpha\dot\beta}
{\delta^{i}}_{k}.
\end{array}
\end{equation}
The rest of the P.B's. for the  supercoordinates $x_{\alpha\dot\alpha},\nu_{a},\theta^{i}_{\alpha}$ coincides with the P.B's. (\ref{66/4'}).

\section{Lorentz invariant brackets with two spinors}

Up to now we have studied the Lorentz invariant  Poisson and Moyal brackets including only one auxiliary  Majorana  spinor $\nu\equiv-\gamma_{5}\lambda$ to construct the scalar derivatives  (\ref{5/32}) from the supersymmetric derivatives $ D^{\alpha},
{\bar D}^{\dot\alpha}$ and 
$\partial^{\alpha\dot\alpha}$ (\ref{3/30}). Using only these  scalars in the Poisson/ Moyal brackets restricts the admissible motions in superspace. To extend the 
set of Lorentz invariant supersymmetric derivatives we need more auxiliary spinors to form the complete spinor basis in $D$-dimensional Minkowski space.
 For  $D=4$ it is enough to add only one new  Majorana spinor $(\mu_{\alpha}, 
\bar\mu_{\dot\alpha})$ forming  the complete spinor basis  together with  
$(\nu_{\alpha}, \bar\nu_{\dot\alpha})$. The  pair  $\nu_{a}, \mu_{a}$ of Majorana spinors  may be identified with the Newman-Penrose dyad \cite{Penrose} if the relations 
\begin{equation}\label{66}
\mu^\alpha \nu_\alpha\equiv \mu^\alpha\varepsilon_{\alpha\beta}\nu^\beta=1,
\quad
\mu_\alpha\nu_\beta - \mu_\beta\nu_\alpha=\varepsilon_{\alpha\beta}
\end{equation}
for the Weyl spinors $\nu_{\alpha},\mu_{\alpha}$ and their complex conjugate are used.
 Having  this spinor basis one can form four real independent Lorentz  invariant supersymmetric differential operators 
\begin{equation}\label{67}
\begin{array}{c}
D^{(\nu)}= \nu_{\alpha}D^{\alpha},
\quad 
\bar D^{(\nu)}= \bar\nu_{\dot\alpha}\bar{D^{\dot\alpha}},
\quad 
D^{(\mu)}= \mu_{\alpha}D^{\alpha},
\quad 
\bar D^{(\mu)}= \bar\mu_{\dot\alpha}\bar{D^{\dot\alpha}},
\end{array}
\end{equation}
two of which $D^{(\nu)},\bar D^{(\nu)}$ coincide with the operators $D,\bar D$ (\ref{5/32}).
 Their linear combinations 
\begin{equation}\label{68}
\begin{array}{c} 
D^{(\nu)}_{\pm}\equiv D^{(\nu)}\pm \bar D^{(\nu)},
\quad 
D^{(\mu)}_{\pm}\equiv D^{(\mu)}\pm \bar D^{(\mu)},
\end{array}
\end{equation}
 form four Lorentz  invariant and supersymmetric  supersubalgebras
\begin{equation}\label{69}
\begin{array}{c} 
[D^{(\nu)}_{\pm}, D^{(\nu)}_{\pm}]_{+}=
\mp8i\partial^{(\nu)},
\quad 
[ D^{(\nu)}_{\pm},  \partial^{(\nu)})]=
[ \partial^{(\nu)},\partial^{(\nu)}]=0,
\quad
\partial^{(\nu)}\equiv(\nu_{\alpha}{\bar\nu}_{\dot\alpha} \partial^{\alpha\dot\alpha}),
\\[0.2cm] 
[D^{(\mu)}_{\pm}, D^{(\mu)}_{\pm}]_{+}=
\mp8i\partial^{(\nu)},
\quad 
[ D^{(\mu)}_{\pm},  \partial^{(\mu)})]=
[ \partial^{(\mu)},\partial^{(\mu)}]=0,
\quad
\partial^{(\mu)}\equiv(\mu_{\alpha}{\bar\mu}_{\dot\alpha} \partial^{\alpha\dot\alpha}),
\end{array}
\end{equation}
which are connected by the anticommutation relations 
\begin{equation}\label{70}
\begin{array}{c}
[D^{(\nu)}_{\pm}, D^{(\mu)}_{\pm}]_{+}=\mp4i \partial^{(+)},
\quad 
\partial^{(+)}\equiv(\nu_{\alpha}{\bar\mu}_{\dot\alpha}+ \mu_{\alpha}{\bar\nu}_{\dot\alpha})
 \partial^{\alpha\dot\alpha},
\\[0.2cm] 
[D^{(\nu)}_{\pm}, D^{(\mu)}_{\mp}]_{+}=\pm4i \partial^{(-)},
\quad 
\partial^{(-)}\equiv(\nu_{\alpha}{\bar\mu}_{\dot\alpha}- \mu_{\alpha}{\bar\nu}_{\dot\alpha})
 \partial^{\alpha\dot\alpha}.
\end{array}
\end{equation}
It is  easy  to see that the Lorentz invariant supersymmetric differential operators  
$D^{(\nu)}_{\pm},  D^{(\mu)}_{\pm},\partial^{(\nu)},
\partial^{(\mu)},\partial^{(\mp)}$ 
describe the whole class of  admissible   motions in the superspace.  These operators together with the extended dilatation operator  $\Delta'$
\begin{equation}\label{71}
\begin{array}{c}
\Delta'=(\nu_{\alpha}
\frac{\partial}{\partial\nu_\alpha} + \bar\nu_{\dot\alpha}\frac {\partial}{\partial\bar\nu_{\dot\alpha}})
-
(\mu_{\alpha}
\frac{\partial}{\partial\mu_\alpha} + \bar\mu_{\dot\alpha}\frac {\partial}{\partial\bar\mu_{\dot\alpha}})
\end{array}
\end{equation}
preserving the condition (\ref{66}) may be used as invariant building blocks for the construction of 
more general Lorentz invariant supersymmetric Poisson and Moyal brackets. 
Using them one can propose the  Lorentz invariant and supersymmetric Poisson  bracket 
\begin{equation}\label{72}
\begin{array}{c} 
\{ F, G \}= F\,[\, -\frac{i}{4}\,(
{\stackrel{\leftarrow}{D}}^{(\nu)}_{-}
{\stackrel{\rightarrow}{D}}^{(\nu)}_{-}+ 
{\stackrel{\leftarrow}{D}}^{(\mu)}_{-}
{\stackrel{\rightarrow}{D}}^{(\mu)}_{-})+
c(\stackrel{\leftarrow}{\partial^{(\nu)}}+\stackrel{\leftarrow}{\partial^{(\mu)}})
\stackrel{\rightarrow}{\Delta'} - 
\stackrel{\leftarrow}{\Delta'}(\stackrel{\rightarrow}{\partial^{(\nu)}}
+\stackrel{\rightarrow}{\partial^{(\mu)}})
\,]\,G .
\end{array}
\end{equation}
 as a generalizations of (\ref{13/31}). The P.B. (\ref{72}) yields the following  coordinate  P.B's.
\begin{equation}\label{73}
\begin{array}{c}
\{x_{m},x_{n}\}=-i(\psi^{(\nu)}_{n}\psi^{(\nu)}_{m} +\psi^{(\mu)}_{n}\psi^{(\mu)}_{m}),
\\[0.2cm] 
\{x_{m},\theta_{a} \}= -\frac{1}{2}(\psi^{(\nu)}_{m}\lambda^{(\nu)}_{a}+ 
\psi^{(\mu)}_{m}\lambda^{(\mu)}_{a}),
\\[0.2cm]
\{\theta_{a},\theta_{b} \}=-\frac{i}{4}(\lambda^{(\nu)}_{a}\lambda^{(\nu)}_{b} + 
\lambda^{(\mu)}_{a}\lambda^{(\mu)}_{b}), 
\end{array}
\end{equation}
where the additional spinor $\lambda^{(\mu)}_{a}$ and grassmannian vector $\psi^{(\mu)}_{m}$ are defined by the relations
\begin{equation}\label{74}
\begin{array}{c}
\quad
\psi^{(\nu)}_{m} \equiv\psi_{n},
\quad 
\lambda^{(\nu)}_{a}\equiv\lambda_{a},
\quad 
\psi^{(\mu)}_{m}\equiv
{1\over 2}(\bar\theta\gamma_{m}\lambda^{(\mu)}),
\quad 
\lambda^{(\mu)}_{a}\equiv(\gamma_{5}\mu)_{a}
\end{array}
\end{equation}
The Majorana spinors $\lambda^{(\nu)}_{a}$ and $\lambda^{(\mu)}_{a}$ have zero P.B's. between themselves and with $\theta_a,\psi^{(\nu)}_{m}, \psi^{(\mu)}_{n}$,  but non zero  P.B's. with  $x_{m}$ 
\begin{equation}\label{75}
\begin{array}{c}
\{x_{m},\lambda^{(\nu)}_{a}\}=c(\varphi^{(\nu)}_{m}+\varphi^{(\mu)}_{m})\lambda^{(\nu)}_{a}),
\quad
\{x_{m},\lambda^{(\mu)}_{a}\}=-c(\varphi^{(\nu)}_{m}+\varphi^{(\mu)}_{m})\lambda^{(\mu)}_{a},
\\[0.2cm] 
\varphi^{(\nu)}_{m}\equiv\varphi_{m},\quad
\varphi^{(\mu)}_{m}\equiv{1\over 4}(\bar\lambda^{(\mu)}\gamma_{m}\lambda^{(\mu)}),
\end{array}
\end{equation}
where the real constant  $c$ has to be defined from the analysis of the Jacobi identities. We see that the addition of new spinors permits to define  more wide class of Lorentz  invariant deformations. 
From the physical point of view these extensions permit to take into account deformations associated with tensile branes or massive fields in the addition to the above considered 
deformations associated with tensionless branes or massless fields. It follows from the results \cite{GuZ},\cite{ZuL}, where the transition from tensionless string/brane \cite{Znull} to tensile one was described in geometrical terms.

\section{Discussion}

We constructed selfconsistent Poisson and Moyal brackets desribing Lorentz invariant supersymmetric deformations of the $N=1$ superspace ($x_m, \theta_a$) equiped by commuting Majorana spinor $\lambda _{a}$ (equivalently $\nu\equiv-\gamma_{5}\lambda$). We proved their selfconsistency and  
 found that the noncommutativity of $x_m$ with $x_n$ or $\theta_a$ is parametrized by composite Ramond vector $\psi_{m}\equiv -\frac{1}{2}(\bar\theta\gamma_{m}\lambda )$ partially accompanied by the grassmannian axial  vector $\psi_{1m}\equiv \frac{1}{2}(\bar\theta\gamma_{m}\gamma_{5}\lambda )$. 
The Ramond vector $\psi_{m}$ originating from the spinninig string/particle models is associated with the spin structure of the enlarged superspace.
The nonanticommutativity of the $\theta_a$ componets among themselves depends only on spin tensors constructed from the auxiliary spinor  which may be treated as a  component of twistor.
It points out that a hidden spinor structure of space-time associated with the Penrose twistor picture could underly the  non(anti)commutativity.
We found a one to one correspondence between the Lorentz invariant Moyal brackets (\ref{48/43}) of the supecoordinates  and  their  background  dependent (anti)commutators parametrized by  the antisymmetric field $B_{mn}$, the graviphoton $C_{ab}$ and the gravitino $\Psi^{a}_{m}$. The map was schematically presented as 
\begin{equation}\label{76}
\begin{array}{c}
B^{-1}_{mn}\leftrightarrow i\psi_{m}\psi_{n},\quad
C_{ab}\leftrightarrow \lambda_{a}\lambda_{b},\quad
 \Psi^{a}_{m}\leftrightarrow \psi_{m}\lambda^{a}
 \end{array}
\end{equation}
and is valid up to the second order corrections in the deformation parameter $h$.
 The map (\ref{76}) transforms the background  dependent  and Lorentz noninvariant (anti)commutators of the  supercoordinates into their invariant Moyal brackets (\ref{48/43}) restoring  desirable Lorentz invariance of the deformations. The map gets a natural explanation in the frame of the Feynman-Wheeler action at-a-distance theory and its superymmeric generalization \cite{TZ}, where the (super)fields were constructed from the (super)space coordinates resulting in  the Maxwell and  Dirac equations.
 We studied the null twistor realization of the brackets and observed the dependence of the non(anti)commutativity effect on the choice of supercoordinate variables. We found that transition to the null supertwistor subspace \cite{Ferb} washes off the noncommutativity effect among the null supertwistor components, because of the nonlocal connection between (super)twistors and (super)coordinates. It shows that the non(anti)commutativity of the original  coordinates of the superspace may be hidden in their nonlinear combinations describing supersymmetric (anti)commutative hypersurfaces embedded in the primary supespace. This observation attracts an attention to the paper \cite{BeZ}, where the new super (D)p-brane models with the  $OSp(1,2M)$ spontaneously broken symmetry were constructed using supertwistor space.
We outlined  some  generalizations of the studied invariant brackets to the cases of $N$ extended supersymmetry and additional spinor coordinates based on the possibility to construct additional Lorentz invariant supersymmetric derivatives. 
An attractive  feature of the  deformation defined by the Moyal brackets originated  from 
 the chiral Poisson bracket (\ref{66'}) is the appearance of composite axial vector $\psi_{1m}={1\over 2}(\bar\theta\gamma_{m}\gamma_{5}\lambda)$ in the pair with Ramond vector  $\psi_{m}$. 
It remindes on $V-A$ structure of the chiral sigma models, electroweak interactions and parity breaking  and  sends  a signal to think about non(anti)commutative deformations of underlying  superspaces as a geometrical source behind the physics of the Standard Model. 
The  microscopic scale of the superspace deformations is fixed by the above introduced deformation parameter $h$. We suppose that effect  of the proposed  deformations on the structure of minimal supersymmetric SM  deserves an attention. The same  concerns a  possible role of the  deformations in the problem of supersymmetric dark matter and dark energy. The work is in progress.

\section{Acknowledgements}

The author thanks Fysikum at the Stockholm University for kind hospitality and I. Bengtsson, L. Bergstr\"om, M. Berg, S. Hannestad and H. Nielsen for useful discussions. The work was partially supported by the grant of the Royal Swedish Academy of Sciences and by the SFFR of Ukraine under Project 02.07/276. It is a pleasure to thank the organizers of the Conference ``Cosmology 2005: a reality check'' in  K{\o}benhavn  for the kind hospitality.


\begin{thebibliography}{99}

\bibitem{Sny}
H.S. Snyder,  Phys. Rev.
{\bf 71} (1947) 38 . 
\bibitem{Casal}
R. Casalbuoni,  
Nuovo. Cim.  {\bf A 33}  (1976) 389. 
\bibitem{BFFLS}
F. Bayen, M. Flato, C. Fronsdal, A. Lichnerowicz and D. Sternheimer,\\
Ann. Phys. {\bf 111} (1978) 61; 111.
\bibitem{BrSch}
L. Brink and J.H. Schwarz,   
Phys. Lett. {\bf B 100} (1981) 310.
\bibitem{SchNwh}
 J.H. Schwarz and  P. van Nieuwenhuizen,
Lett. Nuovo. Cim.  {\bf 34}  (1982) 21.
\bibitem{Con}
A. Connes, 
\emph{Noncommutative  Geometry }, Academic Press, London, 1990. 
\bibitem{Man}
Yu.I. Manin, 
Comm. Math. Phys. {\bf B 123}  (1989) 163.
\bibitem{Kon}
M. Kontsevich, 
arXiv:q-alg/9709040.
\bibitem{BaFiShSu}
T. Banks, W. Fischler, S.H.  Shenker and L. Susskind,  
Phys. Rev. {\bf D 55}  (1997) 5112; arXiv:hep-th/9610043.
\bibitem{CoDoSch}
A. Connes, M. Douglas and A. Schwarz,  
JHEP {\bf 02} (1998) 003; arXiv: hep-th/9711162.
\bibitem{SeWi}
N. Seiberg and E. Witten,  
JHEP  {\bf 09} (1998) 032; arXiv: hep-th/9908142.
\bibitem{MaSSWes}
J. Madore, S. Schraml, P. Schupp and J. Wess, 
  Eur. Phys. J. {\bf C 16} (2000); arXiv:  hep-th/0001203.
\bibitem{KLMa}
P. Kosinski, J. Lukierski and P. Maslanka,
 arXiv: hep-th/00112053.
\bibitem{DoN} 
 M. Douglas and  N. Nekrasov,   
Rev. Mod. Phys.  {\bf 73} (2001) 977; 
arXiv:hep-th/0106048.
\bibitem{Sza}
R.J. Szabo,  
arXiv:hep-th/0109162.
\bibitem{FeL}
S. Ferrara and  M.A. Lledo,  
JHEP {\bf 05} (2000) 008;  arXiv: hep-th/0002084;\\
S. Ferrara, M.A. Lledo and O. Macia, 
 JHEP {\bf 09} (2003) 068;  arXiv: hep-th/0307039.
\bibitem{OoVa}
H. Ooguri and C. Vafa, 
Adv. Theor. Math.Phys.
{\bf 7} (2003) 53; arXiv: hep-th/0104190;  
arXiv:hep-th/0303063;
\bibitem{KPT}
D. Klemm, S. Penati and L. Tamassia,  
Class.Quant.Grav. {\bf 20}
 (2003) 2905; arXiv: hep-th/0104190.
\bibitem{BoGNwh}
J. de Boer, P.A. Grassi and P. van Nieuwenhuizen,
Phys. Lett. {\bf B574} (2003) 98;
 arXiv: hep-th/0302078.
\bibitem{Sei} 
N. Seiberg, JHEP {\bf 06} (2003) 010; arXiv:hep-th/0305248;\\ 
 N. Berkovits and N. Seiberg,  
JHEP {\bf 07} (2003) 010;  arXiv: hep-th/0306226.
\bibitem{ChKNT}
M. Chaichian, P.P. Kulish, K. Nishijima and A. Tureanu, 
 Phys. Lett. {\bf B 604} (2004) 98; arXiv: hep-th/0408069; M. Chaichian, P. Presnajder and A. Tureanu, 
Phys. Rev. Lett. {\bf 94} (2005) 151602; arXiv:hep-th/0408069.
\bibitem{KoSa}
Y. Kobayashi and S. Sasaki, 
arXiv: hep-th/0410164.
\bibitem{BZ}
B. M. Zupnik,  
arXiv: 
hep-th/0506043.
\bibitem{IhM}
 M. Ihl and C. S$\ddot{a}$mann, 
arXiv:  hep-th/0506057. 
 \bibitem{UZnc}
D.V. Uvarov and A.A. Zheltukhin,  
JHEP {\bf 03} (2004) 063; arXiv: hep-th/0310284;\\
 Mod. Phys. Lett. {\bf A 20 } (2005) 769
\bibitem{ZUm}
 A.A. Zheltukhin  and D.V. Uvarov,  
Phys. Lett. {\bf B 545} (2002) 183; 
\bibitem{Penrose}
R. Penrose and M.A.H. McCallum, 
 Phys. Rept. {\bf 6} (1972) 241.
\bibitem{Ferb} 
A. Ferber, 
Nucl. Phys. {\bf B 132} (1978) 55.
\bibitem{Witt}
E. Witten, 
 Phys. Lett. {\bf B77} (1978) 215;
Commun. Math. Phys. {\bf 252} (2004) 189;  arXiv: hep-th/0310284.
\bibitem{Shir}
T. Shirafuji,  
Prog. Theor. Phys.
 {\bf 70} (1983) 18.
\bibitem{BC}
I.  Bengtsson and M. Cederwal,
Nucl. Phys. {\bf B 302} (1988) 81.
\bibitem{GSW}
M.B. Green, J.H. Schwarz and E. Witten, 
\emph{Superstring theory} Cambridge University Press,
Cambridge 1987. 
\bibitem{BDZDH}
L. Brink, P. Di Vechia, P. Howe and B. Zumino,
 Phys. Lett. {\bf B 64} (1976) 435.
\bibitem{Zkk}
 A.A. Zheltukhin,  
Phys. Lett. {\bf B 168} (1986) 43; 
\bibitem{VZ}
D.V. Volkov and  A.A. Zheltukhin, 
JETP Lett. {\bf 48} (1988) 63;
Lett. Math. Phys. {\bf 17} (1989) 141; 
 Nucl. Phys. {\bf B 335} (1990) 723. 
\bibitem{BW}
J. Wess and J. Bagger,\emph{ Supersymmetry and Supergravity}, Princeton, USA, Univ. Press (1992).
\bibitem{Zmoy}
 A.A. Zheltukhin,  arXiv: hep-th/0506127.
\bibitem{TZ}
V.V. Tugai and A.A. Zheltukhin,
Phys. Rev. {\bf D 51} (1995) 39; ibid.  {\bf D 54} (1996) 4160; 
Phys. of Atomic Nucl. {\bf 61} (1998) 274.
\bibitem{Znull}  
 A.A. Zheltukhin, Sov. J. Nucl. Phys.  {\bf 48} (1998) 375.
\bibitem{GuZ}
O.E. Gusev and  A.A. Zheltukhin, JETP Lett.  {\bf 64} (1996) 487.
\bibitem{ZuL}  
 A.A. Zheltukhin and U. Lindstr\"om, Nucl. Phys. (Proc. Suppl.) {\bf 102}  (2001) 126;\\
J. High Energy Phys. {\bf 01}  (2002) 034.
\bibitem{BeZ}
I. Bengtsson and A.A. Zheltukhin,  Phys. Lett. {\bf B 570} (2003) 222.

\end{thebibliography}
\end{document}